# Radio frequency for particle accelerators – evolution and anatomy of a technology

*M. Vretenar*
CERN, Geneva, Switzerland

**Abstract**
This introductory lecture outlines the impressive progress of radio frequency technology, from the first table-top equipment to the present gigantic installations. The outcome of 83 years of evolution is subsequently submitted to an anatomical analysis, which allows identifying the main components of a modern RF system and their interrelations.

## 1   From Maxwell to the radio

We can incontestably assert that modern electromagnetism starts with the publication of the Maxwell equations in 1864. The list of practical applications of Maxwell's theory is endless, but here in particular we are interested in the development of the radio, which is going to be the starting point for RF technology.

We know that Maxwell himself developed from his equations the theoretical basis for wave propagation in the years around 1873, however, it was only in 1888 that Heinrich Hertz was able to generate (and detect) electromagnetic waves. From this moment, progress was extremely rapid, and in 1891 Nikola Tesla and Guglielmo Marconi laid the basis for what at the time was called the 'wireless telegraph'. The wireless was extremely simple, very far from modern radio, the goal being just to create with a spark gap strong pulses with a wide frequency spectrum that could be detected at long distances. It was only the invention of vacuum tubes in the early twentieth century that made possible the transmission of single-frequency RF waves. In particular, the triode was patented by De Forest in 1907, although at the time he did not have in mind any specific application for his invention. It was only during World War I and because of the war needs that tubes started to be used for communications, with the positive consequence that their cost went down because of the large production quantities and that several technological improvements were introduced, like cathode coatings for improved emission.

The technology was now mature for the invention of the radio, and in the years immediately following the war several teams started to develop broadcasting systems, based on the technology of diffusing from an antenna high-power fixed frequency waves, modulated in amplitude by a low-frequency audio signal. The first radio broadcasting in history came from a Dutch amateur in 1919; during the 1920s all major countries started official radio broadcasting, the first being Argentina in 1920, immediately followed by the US and most European countries.

The result was that from about 1925 all the technology required for radio broadcasting was available at relatively low cost: vacuum tubes (triodes) able to produce high-power (for the time) waves at a frequency up to few MHz, with all related electronics: oscillating circuits, components, etc.

## 2   From Rutherford to the particle accelerator

The years immediately following WWI were particularly exciting not only for technology, but also for basic science. After much theoretical work during the first years of the century aimed at understanding the structure of the atom, immediately after the war atomic physics started to become an experimental science. The milestone in this respect was the historical experiment by Ernest Rutherford, who in 1919 announced his success in producing the first disintegration of the nucleus. With a very simple apparatus, reproduced in Fig. 1, he bombarded nitrogen gas with α-particles generated by radioactive decay of radium and thorium, producing a small amount of hydrogen from the splitting of the nitrogen nuclei. This experiment made a tremendous impact on the general public: scientists had realised the old dream of medieval alchemists, transform the elements one into another.

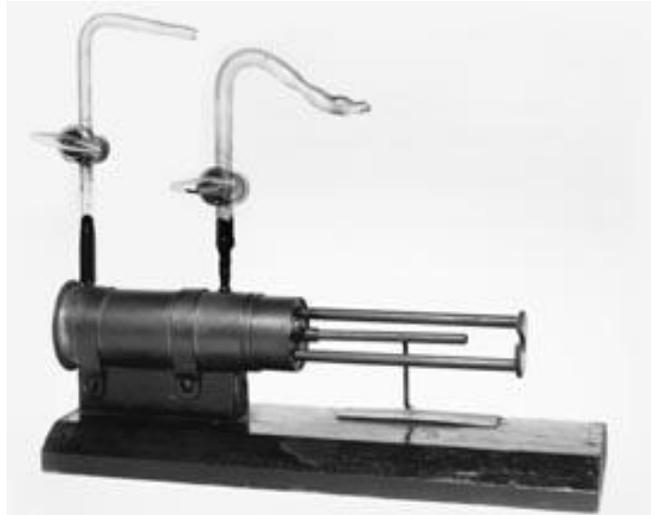

**Fig. 1:** Reproduction of the Rutherford chamber

The main consequence of Rutherford's experiment was more and more interest in experimental nuclear physics. It was immediately clear that the main limitation for nuclear experiments was the projectile: with the particles coming from radioactive decays, only a few light atoms could be modified, and in insignificant quantities. In this respect, there was a large resonance from the plea by Rutherford at a speech at the Royal Society in 1927 [1], where he asked the scientific community to start developing devices that could supply large amounts of charged particles at the energy required to disintegrate the nuclei. This was in fact a major technological challenge: the value of the Coulomb barrier formed by the electrons around the nucleus was estimated at about 500 kV, and to be useful 'particle accelerators' had to reach energies beyond this threshold.

Rutherford's plea did not go unnoticed, and the main laboratories in Europe and in the US started to develop accelerators based on high-voltage technology, i.e., 'electrostatic' accelerators. Many possible ways of reaching very high voltages were explored at the time: Cockcroft and Walton at Cambridge started to develop their well-known high-voltage generator; Van de Graaf at Princeton worked at a belt-charged generator; and other teams around the world explored different approaches, like pulsed techniques, capacitor discharges, transformers, etc.

Eventually, the winners of the race to high voltage and high energy were Cockcroft and Walton, who in 1932 announced that they had obtained disintegration of lithium by 400 keV protons (the theoreticians had been too pessimistic); Fig. 2 is a famous photograph of Walton inside the electronic cage of his generator. However, it appeared immediately that high-voltage technology applied to accelerators could not be improved further: electrostatic accelerators were already hitting the hard limit represented by discharge between the electrodes, whereas disintegration of heavier nuclei in large quantities required higher and higher energies. A new technology was needed to bring accelerators out of the dead end.

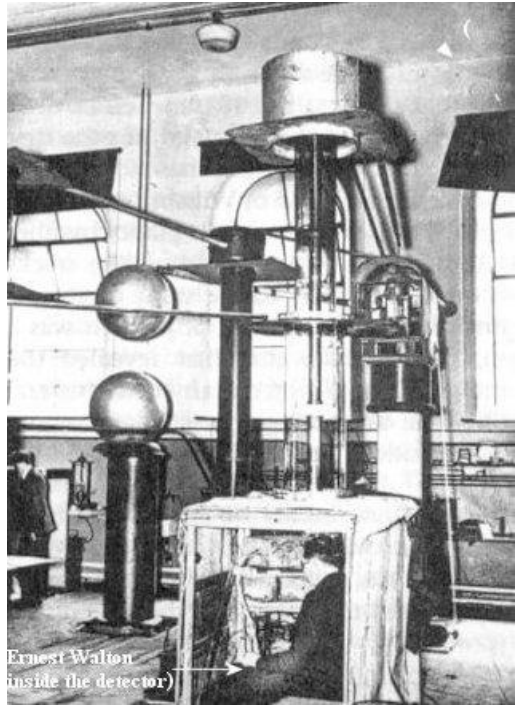

**Fig. 2:** The first Cockcroft–Walton electrostatic generator (with E. Walton inside the detector)

## 3   Marrying the radio and the particle accelerator: Wideröe and Lawrence

We have seen that in the second half of the 1920s radio technology was already mature and widespread, and on the other hand that there was an enormous interest in technologies for the acceleration of particles. It is clear that a breakthrough could come from the marriage of radio technology with particle accelerators, but who was the first to have the idea?

Surprisingly (or maybe not) the idea came from a young PhD student. Rolf Wideröe, a Norwegian student of electrical engineering at Aachen University in Germany, not only was the first to propose the principle of radio frequency acceleration, but was also able to assemble a table-top experimental RF accelerator which is a masterpiece of ingenuity and practical ability [2]. However, this did not come out in a straightforward way; initially, Wideröe's main interest was the acceleration of intense electron beams for X-ray production, at the time another increasingly important application of accelerators. But unfortunately the X-ray transformer, an ancestor of the betatron, that he built for his thesis did not work correctly (later it turned out that it was because of a construction flaw and not of a design problem) and his professor invited him to find quickly another subject. Wideröe had little time left before the term for the thesis, and willing to realise an experimental device he remembered that in 1924 a Swedish professor, Gustav Ising, had published a paper proposing to accelerate particles between 'drift tubes' excited by pulses of travelling waves [3]. The main idea of Wideröe was to use the recent radio technology to produce an accelerating voltage between the tubes, and to apply it continuously rather than in pulses. But this led to another problem: because the voltage was changing sign with the period of the RF wave, only particles close to the crest could be accelerated. In order to accumulate energy over several drift tubes, the length of the tubes had to match the velocity of the particles and only a fraction of the beam could gain energy. This was the beginning of synchronous acceleration.

The Wideröe experimental device is shown in Fig. 3, taken from his thesis presented to the University in 1927. In this simple table-top equipment he applied to a drift tube 25 kV of RF at 1 MHz and successfully accelerated a few single-charged potassium ions up to 50 keV, demonstrating that contributions from two correctly spaced gaps could be added. It is interesting to observe how this is a real miniature accelerator made of an ion source followed by a double-gap accelerating section fed by an RF system. The entire device is inside a vacuum system and the accelerator is followed by a simple spectrometer.

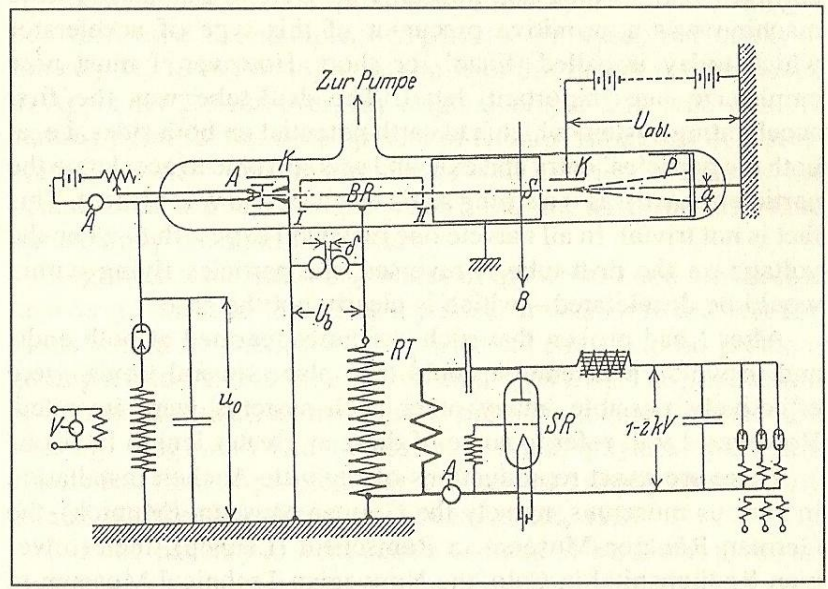

**Fig. 3:** Wideröe's first RF accelerator [2]

Although extremely clever, Wideröe's device did not have a follow-up. With the relatively low frequencies available at the time, it was not possible to imagine the acceleration of protons because the drift tubes would have been excessively long: 5 metres for 500 keV at 1 MHz. For this reason Wideröe published his thesis in the "Archiv für Elektrotechnik" [4] and did not consider pushing further his idea. He eventually went to work for AEG to build high-voltage circuit breakers, and only several years later did he resume developing his betatron, rapidly becoming the main European expert in this field [2]. Years later, Wideröe was an advisor to CERN and can be counted among its founding fathers.

To go one step further from Wideröe's device something new was needed, and this came from the other side of the world, from California. In the 1920s, Ernest O. Lawrence (born in 1901) was a brilliant young professor of physics at the University of Berkeley. He gathered an excellent team of collaborators at his at the time remote University and was frantically looking for a new idea to join the 'energy race'. He himself recalled some years later that in 1929, during a boring conference, he went to look at recent publications at the library, where he spotted the "Archiv für Elektrotechnik" issue with Wideröe's thesis. By all accounts Lawrence did not speak German, but the pictures were so self-explanatory that he immediately realised that radio-frequency acceleration could put Berkeley into the accelerator race. Without delay, he assigned two of his best PhD students to work on two ideas. David H. Sloan started building a Wideröe-type linac for heavy ions made of several drift tubes [5], whereas M. Stanley Livingston was launched on a more exotic idea: a 'cyclic' accelerator bending particles on a circular path around a Wideröe-type drift tube. The latter eventually became the cyclotron, the first real RF accelerator [6]. In the final design, the drift tube is replaced by two hollow D-shaped electrodes; the particles are kept in a circular orbit by the magnetic field generated by a large magnet housing the D's and are accelerated in the gap between the D's. The path inside the D's can be long enough for the particles to remain synchronous with the RF, even at the low frequencies available at the time (3.5 MHz for the first Lawrence cyclotron). Moreover, in a cyclotron at non-relativistic energy the revolution frequency does not depend on the particle velocity and the frequency can be kept conveniently constant.

Figure 4 is another famous picture from a PhD thesis, this time from Livingston's, which shows the basic scheme of the first cyclotron. Again, one of the key components is a triode-based RF system (called oscillator in Livingston's scheme). Lawrence's team succeeded in 1931 to reach a record 1.2 MeV with its first cyclotron, achieving in 1932 the first nuclear disintegrations. A larger cyclotron was able to go as high as 5 MeV in 1935, paving the way for the extensive use of radio frequency in particle accelerators.

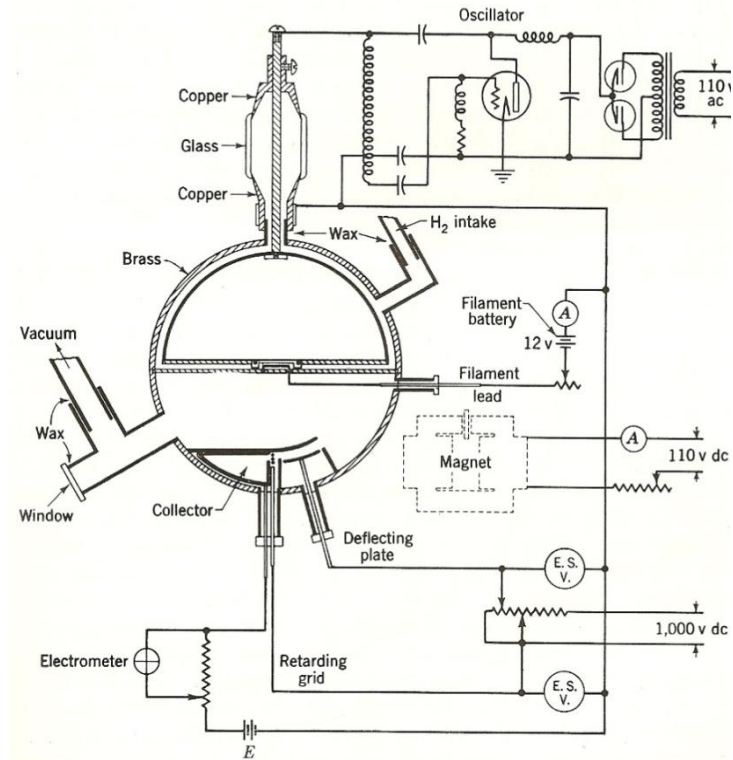

**Fig. 4:** Scheme of Lawrence and Livingston's first cyclotron [7]

## 4 Towards higher frequencies: cavities, klystrons, and the radar

The invention of the cyclotron definitely placed radio frequency at the centre of particle accelerators; however, it also showed that more progress was needed in this field if one wanted to go beyond the limitations of Lawrence-type cyclotrons. In order to reach higher energies and keeping the dimensions reasonable, higher magnetic fields but also higher RF frequencies were needed. But trying to push standard radio technology based on oscillating circuits towards higher frequencies was forbidden by the limitation coming from the power radiated from the circuit, which appears as soon as the RF wavelength becomes comparable with the circuit dimensions. To go to higher frequencies a new approach to building resonant circuits was needed; eventually, while working on the linear accelerator with Sloan at Berkeley, William W. Hansen was the first to propose using a new concept, the 'cavity resonator'.

When Hansen moved to Stanford University in 1937 he continued his research on resonant cavities there, laying the foundations for modern microwave technology. In this period, from his fortunate collaboration with the brothers Russell and Sigurd Varian came another milestone, the invention of the klystron [8]. The early klystrons were low-power devices, which used the new cavity technology to produce mW-level power at frequencies at the time unheard of, in the GHz range. The first application of the klystron was in the radars used during WWII (Fig. 5), and it was only after the war that the Varian brothers developed the high-power klystron; in 1948 they left Stanford to open their company to commercialize a wide range of klystrons to be used in the rapidly developing TV broadcasting market, marking the first example of the transfer of a technology from RF for accelerators to industry.

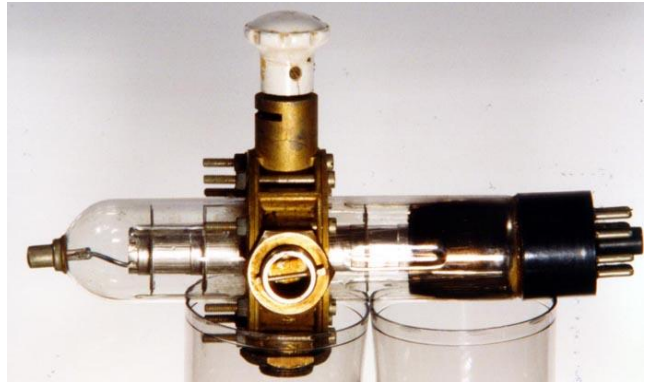

**Fig. 5:** A WWII 3 GHz klystron

However, what definitely brought high-frequency RF out of the US West Coast universities and transformed it into a mature technology with a wide range of applications was the war effort related to radar. While most of the developed countries had a radar programme in the 1930s, the development of modern powerful radar devices is definitely related to the work on high-power magnetron oscillators which took place at Birmingham in the UK during the early 1940s. Lacking the money to develop the magnetron on a large scale, the British government decided to share this strategic war technology with the US at the end of 1940. This resulted in the start of a massive research effort in the US for the development of powerful radar devices to be used during the war. Important units like the famous Radiation Laboratory at MIT were created to work on microwave technology; they immediately enrolled the best accelerator scientists, who were the only people with experience in high frequencies. The result was impressive and during the few years of the war these teams developed all of modern microwave technology: waveguide theory, coupling techniques, microwave measurement equipment and techniques, etc. After the war, the US government realised the industrial potential of these technologies, and decided to declassify the results achieved by radar-related research. This resulted in the publication of the well-known series of RF books and in a much larger use of RF technology, in the broadcasting and radar industry but also in the particle accelerator field. Scientists finally had the technology to go up in frequency and make major improvements on the first generation of accelerators.

As an example of a scientist fighting the radar war, we can quote the interesting case of Hans Bethe and coupling. Bethe was a famous theoretical physicist and nuclear scientist, who at the rise of Nazism left Germany for the UK and then the US. When the war started, he volunteered to contribute to the war effort, and was put in touch with the MIT radar team. After wondering about what to ask a theoretical physicist, they decided to propose to him to study the problem of coupling via a slot from a waveguide to a cavity and between two cavities. The resulting paper [9] brilliantly solves the problem starting from black-body radiation theory and remains the starting point for all those who after Bethe wanted to study this subject. Immediately after writing this paper Bethe moved to the Manhattan Project, leaving it as his only contribution to radio-frequency technology.

## 5  The first linear accelerators

After the war, accelerator research restarted in earnest, driven by the wide interest in atomic physics related to the development of nuclear weapons and nuclear technology. The accelerator scientists now had both the competences acquired during the war years and the access to low-cost RF technology developed by industry for the war effort. The goal was now to use modern high-frequency technology to bring beams of protons or electrons to unprecedented energies.

The first to move in this direction was Luis Alvarez, a talented experimental physicist from the Berkeley group, who developed with his team the first high-energy proton linear accelerator, the Drift Tube Linac (DTL). What he did was basically placing a Sloan-type drift-tube structure inside a large Hansen-type cavity resonator. Because of the operating mode the accelerating field sequence in a DTL is different than in a Wideröe-type structure, requiring the individual cells (drift tube and gap) to be $\beta\lambda$ long for the particle to remain synchronous with the field; the consequence is that in order to accelerate protons up to energies of the order of few tens of MeV ($\beta \sim 0.1$–$0.2$) with drift tubes of reasonable length, the RF frequency must be higher than about 100 MHz, a frequency range now accessible thanks to the development of radar technology.

The first DTL built by Alvarez had an even closer contact with radar: the frequency of 202.56 MHz was chosen because the US Army made available to him a stock of surplus radar amplifiers at this frequency, to be used for his new linac. Eventually, Alvarez became the owner of 2000 units of 85 kW each, an impressive stockpile, out of which he used only 26 units for his 32 MeV linear accelerator. Figure 6 shows the first Alvarez DTL with the RF amplifiers on both sides. In 1948 this first linac reached the design energy, paving the way for a new generation of high-energy accelerators [10]. From that moment the 202.56 MHz frequency remained a standard for proton linacs, as is the case for the CERN linacs.

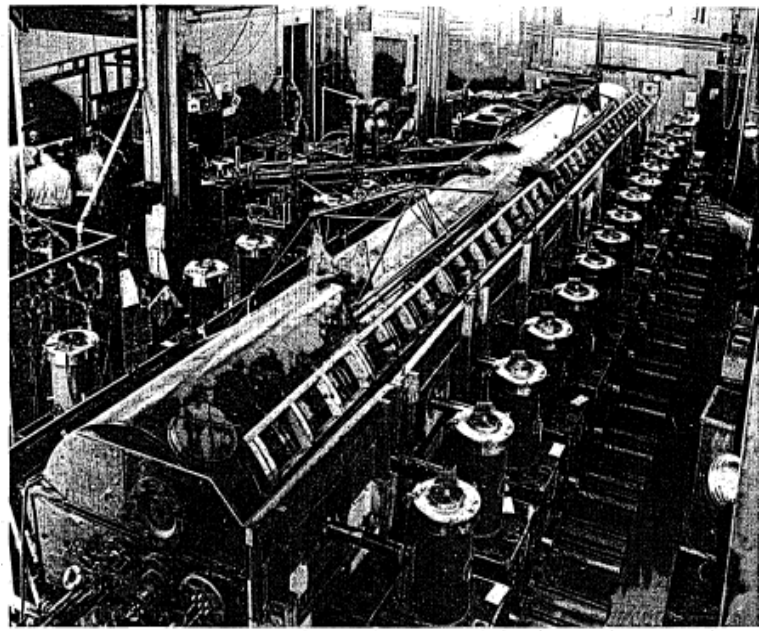

**Fig. 6:** The Alvarez drift tube linac [10]

At about the same time, the Stanford team was also progressing towards high frequencies, but in this case for electrons, and out of the work of a first-class team made by Hansen, Ginzton and Kennedy came the principle of the 3 GHz electron linac, based on travelling-wave, iris-loaded structures. Also here the frequency of 3 GHz was chosen because a radar-type magnetron was used as RF power source on their first prototype. The klystrons that were later developed at Stanford kept this frequency, which is still the standard frequency for electron linacs.

Figure 7 shows Hansen and his colleagues with a section of the first electron linear accelerator that operated at Stanford University in 1947. It was 3.6 metres long and could accelerate electrons to 6 MeV. This photograph is famous because it was the Stanford answer to a photo (Fig. 8) published by Berkeley a few months earlier, which shows the first tank of the Alvarez DTL with all the team on top of it. There was no better way to advertise the advantages of higher frequencies!

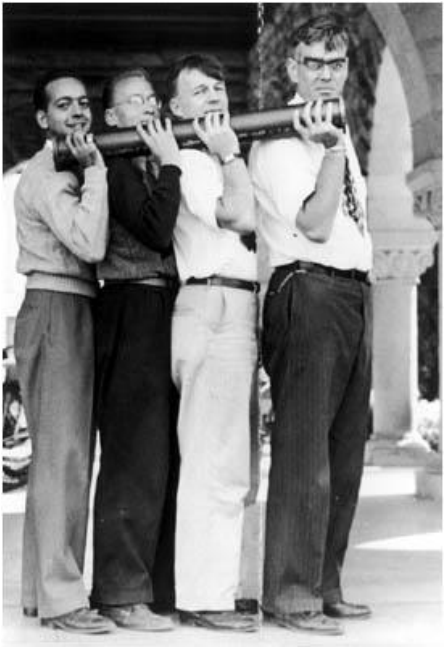

**Fig. 7:** W. Hansen (right) and colleagues with a section of the first Stanford electron linac [10]

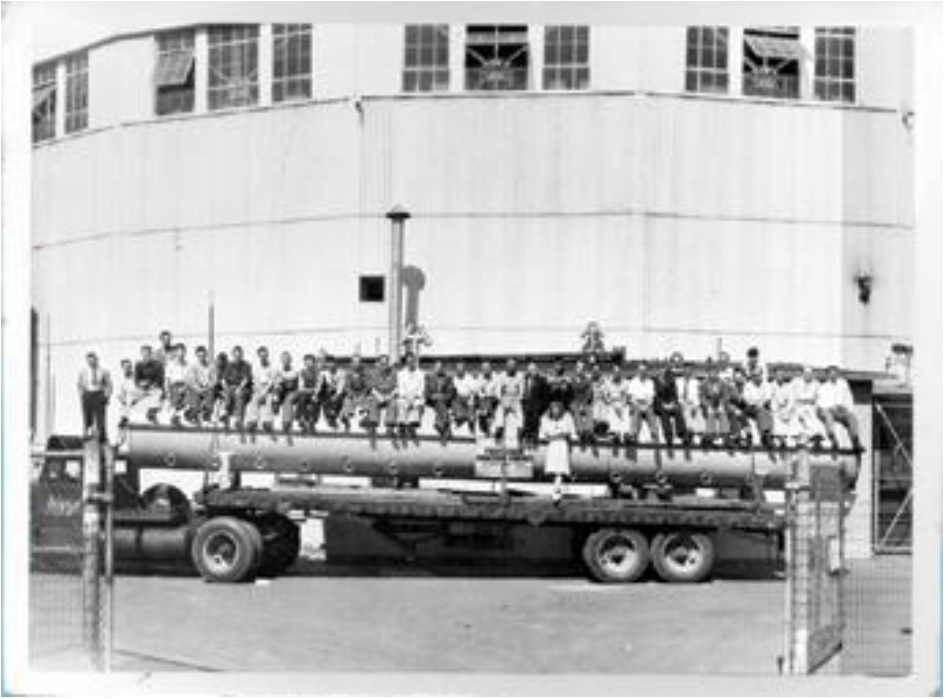

**Fig. 8:** Berkeley Laboratory group seated on top of the vacuum tank of their proton linac [10]

## 6     New challenges: the synchrotron

After the success of the first generation of linear accelerators, the quest for higher energy continued leading to the development of machines capable of bringing protons up to relativistic energies. Early prototype synchrotrons were tried immediately after the war, but only from the early 1950s were large-scale synchrotrons designed and built in the US and UK, reaching energies in the GeV range and introducing new challenges for RF systems.

In a synchrotron the energy is transferred to the beam by only a few elements along the ring, the accelerating cavities. Synchrotron cavities can have a lower frequency than linacs, the equivalent 'drift tube' being the entire accelerator circumference, and a smaller gap voltage because the acceleration takes place over several turns. However, their main specific characteristic is that their resonant frequency has to change during the acceleration period in order to keep the beam of increasing velocity synchronous with the RF fields, something that required the development of a new and sophisticated technology.

After some attempts at different techniques, the introduction of ferrites in the inductive part of the cavity finally solved the problem. An additional difficulty comes from the fact that the ferrites must stay in air, leading to the development for the early synchrotrons of ceramic gap insulators that allow separating the vacuum of the beam pipe from the air of the cavity. Moreover, the need to produce long RF pulses marked the separation from the short-pulse radar technology and led to the development of specific tube-based RF amplifiers. Figure 9 shows the scheme of one of the first modern synchrotron cavities, from the Brookhaven Cosmotron (1952, 3 GeV proton beam energy).

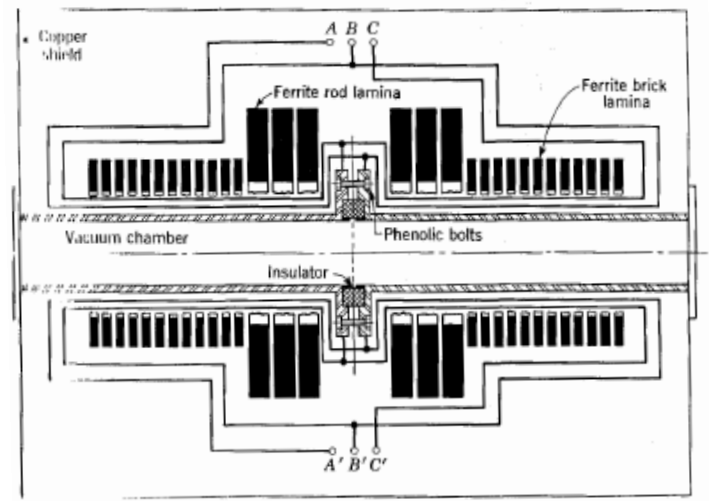

**Fig. 9:** Scheme of the RF cavity for the Brookhaven Cosmotron

An additional challenge for the RF systems of early synchrotrons was the need for a number of RF feedback loops that could keep the voltage and the frequency programmes under control during acceleration. A pioneer in this respect was the CERN PS, commissioned in 1959. The importance of the RF controls loops was demonstrated by the fact that only the introduction of a radial loop hastily assembled in a Nescafe tin by Wolfgang Schnell, the RF group leader, allowed transition to be passed on the night of 24.11.1959 and the beam to be accelerated to full energy after some months of fruitless attempts. Figure 10 is a famous photograph from that night, showing Schnell in front of his oscilloscope surrounded by a cheery project team.

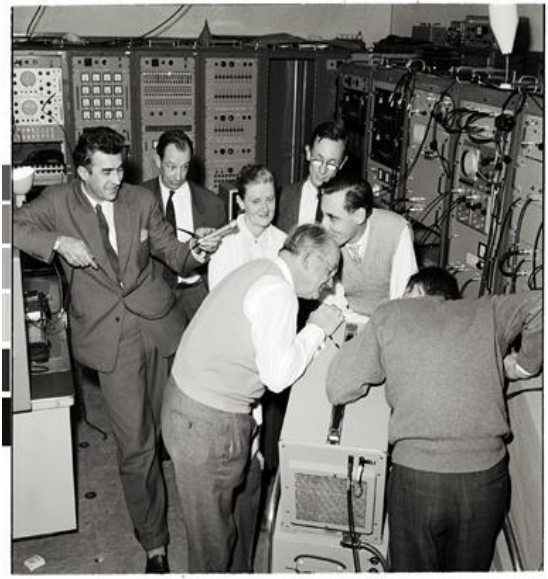

**Fig. 10:** During the first hours of the start-up of the CERN PS (24.11.1959)

## 7  The maturity: high frequency, high power, low loss

The start of the CERN PS marked the transition to the 1960s and to the maturity of RF and particle accelerators. Most of the required technologies had been developed, and what remained was to improve and to exploit them on a large scale. The last 50 years have seen a burgeoning of accelerators around the world, each with its own specific RF system, going from the large colliders for physics to the small machines for industrial applications and medicine: more than 7000 linear electron accelerators for X-ray therapy are operational around the world.

In very general terms, the main trends of the last years have been:

i.  increase in the frequency of the RF systems, in particular for linear accelerators, going up to the 30 GHz of the original CLIC proposal;

ii.  increase in complexity, in particular for the number and quality of control loops;

iii.  increase in RF power (pulsed or CW);

iv.  improvement of the cavity design and construction techniques.

In parallel with the steady improvement of the technologies, only two major innovations appeared in the field of accelerator RF during the last 50 years: superconductivity and computer codes for RF design.

The application of superconductivity to particle accelerators is relatively recent; if superconducting phenomena have been known since 1911, their theoretical understanding based on the BCS theory came only in 1957. A few years later in 1965 the first superconducting accelerating cavity was tested at Stanford, a lead-plated resonator that successfully accelerated a beam of electrons. Immediately accelerator experts were attracted by the idea of freeing themselves from the problem and cost of generating large amounts of RF power, even if the price to pay was operation of the cavity at cryogenic temperatures. During the 1970s several superconducting cavity projects aiming at 2–3 MV/m gradient were launched in different Laboratories (Stanford, Illinois, CERN, Karlsruhe, Cornell, Argonne), using different designs and technologies. Major improvements started to come from the late 1970s and during the 1980s, when improved cleaning techniques plus geometry optimization and improvement in niobium quality allowed one to routinely reach gradients higher than 10 MV/m. This paved the way for the large-scale superconducting RF project that came to life in the 1980s and early 1990s: ATLAS and CEBAF in the USA, HERA and LEP-II (Fig. 11) in Europe. After the successful application of RF superconductivity to large-scale projects, recent years have seen on the one hand impressive developments in the surface preparation technique that allows one to safely operate superconducting RF systems at unprecedented gradients, and on the other hand a reduction in their costs and operational problems that has allowed one to use superconductivity in a wide range of accelerator projects of all sizes.

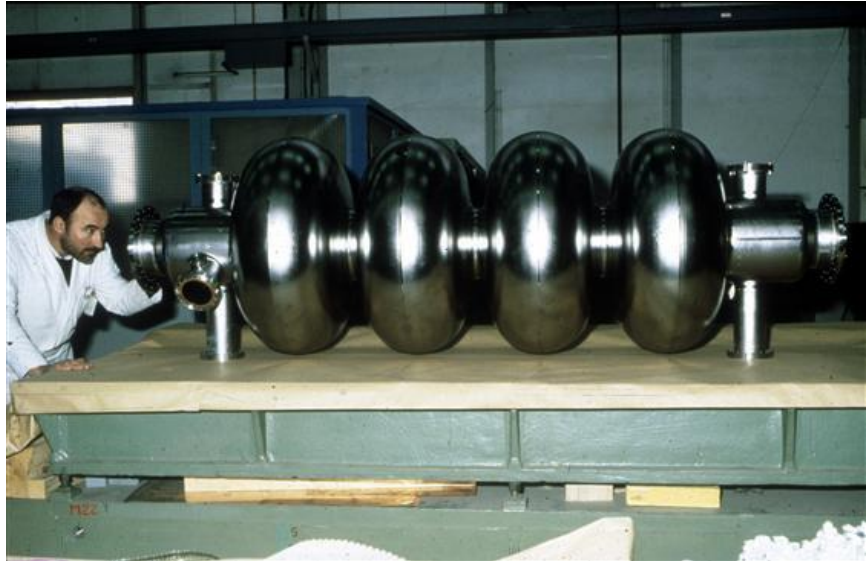

**Fig. 11:** 4-cell prototype superconducting RF cavity for LEP (1985)

Another innovation that deserves to be mentioned is the rise of RF simulation software, which nowadays allows computers to solve the most complicated 3D problems. For years, the only way to define the dimensions of accelerator RF cavities was by scaling from laboratory models, followed by a trial and error optimization. The first rudimentary computer codes for the calculation of small circular-symmetric cavities were developed in the late 1960s, but the first real breakthrough towards computing field distributions and resonance frequencies was the 2-dimensional code Superfish by Klaus Halbach and Ronald Holsinger, which was first presented in 1976 [11]. The first codes were running on large mainframe computers, required long computation times and were limited in the size and definition accuracy of the geometry under study. Over the years, the development of more powerful computers has greatly helped the work of the cavity designer; an important step in this direction was the first PC version of Superfish that appeared in 1985. The next move was the calculation of geometries with a general 3-dimensional shape, and the first code to achieve this goal was MAFIA (solution of MAxwell's equations by the Finite Integration Algorithm), originally developed at DESY under the direction of Thomas Weiland. The first release of the complete MAFIA package took place in 1986 [12].

Nowadays after more than two decades of development a wide number of 3-dimensional RF codes are available on the market, able to solve with extreme accuracy the most complex RF and microwave problems. It is interesting to observe that for the most commonly used 3D codes the design of accelerator cavities represents now a minor fraction of the advertised applications. Among their main applications are now EMC analyses, high-frequency circuit design, antenna design, charged particle dynamics in electromagnetic fields, electro and magneto-static design, etc. The large diffusion of these codes outside of the accelerator field indicates how this has been another successful transfer of technology from accelerator RF to industry.

# 8 Anatomy of an RF system

In order to understand the behaviour of an RF system, it is useful to first analyse its anatomy and then to dismember it into its basic components.

In general terms, an RF system is a very simple device. It can be considered as a black box (Fig. 12) that communicates with the external world through only five channels:

– an input beam and an output beam, with the output beam at higher energy than at input;
– an input for the electrical power coming from the grid;
– an output for the dissipated heat, usually via some cooling water;
– an input/output for the control of the system, which gives to an external operator access to its main parameters.

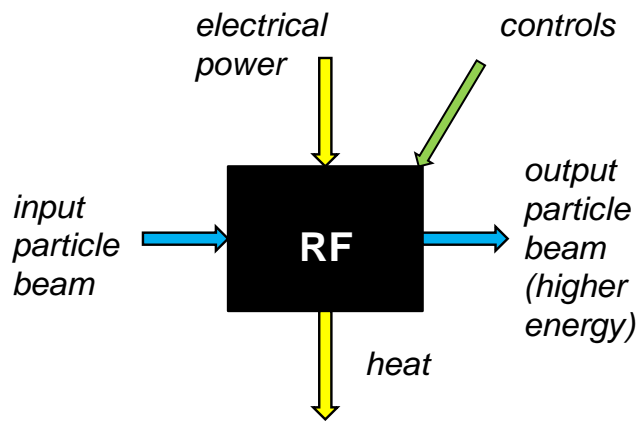

**Fig. 12:** 'Black-box' scheme of an RF system

This representation underlines the fact that an RF system is a device that transforms electrical energy taken from the grid into energy transferred to a beam of particles. The efficiency of this transformation is usually quite low, and a large fraction of the input energy will be dissipated into heat released in the surrounding environment. More precisely, the energy transformation will take place in three different steps, each with its own technology and its own efficiency:

i. The transformation of the AC power from the grid (alternate, low voltage, high current) to DC power (continuous, high voltage, low current) that takes place in a power converter; the efficiency is that of the power converter.

ii. The transformation of the DC power into RF power (high-frequency) that takes place in an RF active element: RF tube, klystron, transistor, etc.; the efficiency is the usual RF conversion efficiency that depends on the specific device and its class of operation.

iii. The transformation of the RF power into power to a particle beam that takes place in the gap of an accelerating cavity; the efficiency is proportional to the shunt impedance, which represents the efficiency of the gap in converting RF power into voltage available for a beam crossing the cavity at a given velocity.

An expanded scheme for an RF system is presented in Fig. 13; it identifies the three main elements of an RF system, power converter, RF amplifier, and RF cavity, together with the control loops (low level RF, including the main oscillator), and the ancillary systems: cooling, vacuum, and cryogenics in the case of superconducting cavities.

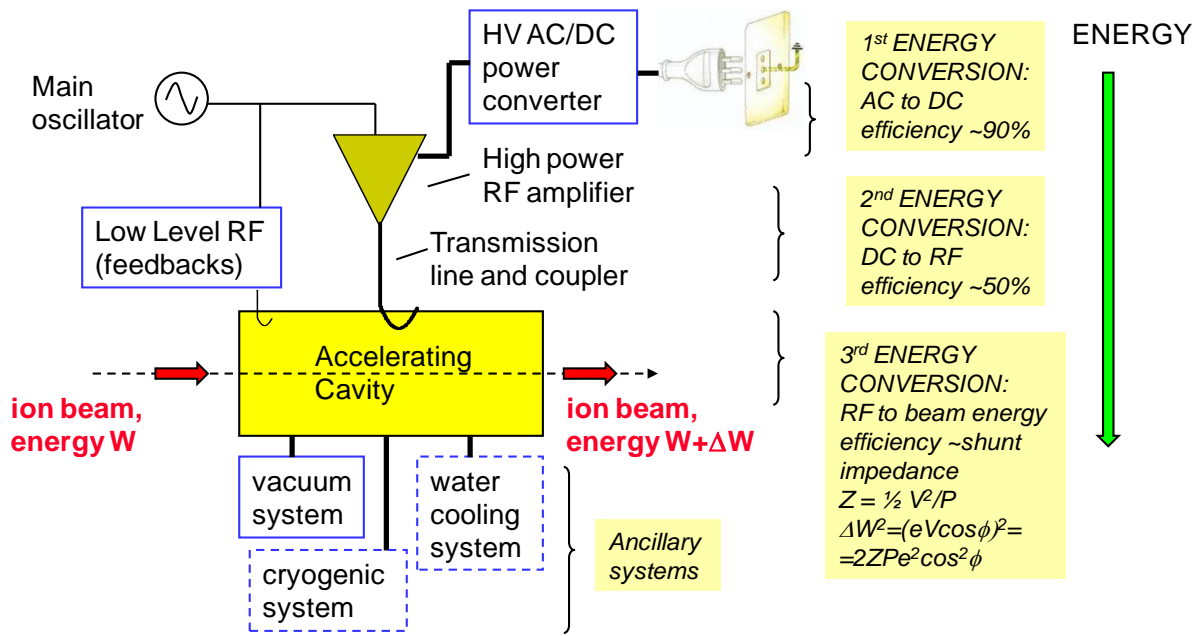

**Fig. 13:** Block-diagram scheme of an accelerator RF system

Looking in detail at the different components, we can identify their main features as:

1. *The power converter:* although it is sometimes subcontracted outside of the RF groups or considered as an integrated part of the RF amplifier, the power converter is an essential part of an RF system not only because it provides the first energy transformation, but because operating with high voltages it is often one of the main items responsible for the overall reliability of an RF system. It usually comprises a rectifying section and a transformer to step up the voltage. Most of the RF systems operate in pulsed mode, and the power converter has to act as a buffer between the pulsed system and the grid, providing a constant charge to the electrical network. This is obtained by storage of the energy in a capacitor bank, either stand-alone or part of a pulse forming network. Pulsed power converters are usually called 'modulators'.

2. *The RF amplifier:* has the important task of converting DC into RF power in an active element. The conversion takes place by means of an electron beam accelerated by the DC voltage and density modulated at the RF frequency by a grid (vacuum tube), by a cavity (klystron), or directly by an applied voltage (transistor). The modulated RF power is then extracted from a resonant cavity excited by the electron beam. The output cavity is either in air around the anode (vacuum tube) or in vacuum around the electron beam path (klystron). Only a fraction of the power of the electron beam goes to the RF; the rest is dissipated in the electron collector.

3. *The transmission line:* the power generated by the RF amplifier has to be transported to the accelerating cavity reliably (without arcing), with minimum loss and zero reflection. Rigid coaxial lines, waveguides, or different types of coaxial cable are commonly used for the transmission lines.

4. *The power coupler:* the power transported to the cavity has to be injected into the cavity by a device that needs at the same time to couple strongly the transmission line to the field distribution in the cavity, to provide a way to adjust the matching between the line and the cavity, and finally to separate the vacuum of the cavity from the air of the line. The consequence of these multiple requirements is that the power coupler is usually a very critical device that requires a careful design and prototyping.

5. *The Low Level RF (LLRF):* it is in a sense the brain of an RF system, because at the same time it concentrates and elaborates the flow of information required to guarantee a stable operation and provides the interface with the operators who need to set up the system parameters. A LLRF is basically a collection of electronics, slow and fast, analog and digital, which controls the return loops from the cavity and/or the beam to the RF amplifiers, required to provide a constant phase and amplitude of the gap RF field in the presence of perturbations coming from the external environment or from the beam. We can consider as part of the LLRF the main oscillator, which generates the reference operating frequency.

6. *The accelerating cavity:* the heart of an RF system, it has to concentrate the RF energy under the form of electric field on a gap, with minimum power loss. Operating in vacuum, it is usually a complicated multidisciplinary object that integrates beam dynamics aspects (sequence and position of gaps, integration of transverse focusing elements), RF design (layout of the RF fields, path and dissipation of the RF currents), mechanical aspects (construction and joining techniques), thermo mechanical aspects (cooling, deformations) and vacuum properties.

**Acknowledgements**

The preparation of this lecture profited from many enlightening discussions with friends and colleagues; to all of them goes my gratitude. In the preparation of the historical introduction an inextinguishable source of information and inspiration has been the recent book on accelerator history by A. M. Sessler and E. J. N. Wilson, *Engines of Discovery: A Century of Particle Accelerators* (World Scientific, 2007).